\documentclass[11pt]{article}
\usepackage[utf8]{inputenc}
\usepackage{graphicx}
\usepackage{amssymb}
\usepackage{amsmath}
\usepackage{amsfonts}
\usepackage{wasysym}
\usepackage{textcomp}
\usepackage{ dsfont }
\usepackage{circuitikz}
\usepackage{csquotes}
\usepackage[english]{babel}
\usepackage{filecontents}

\bibliography{refs}

\usepackage{xcolor}
\usepackage{color} 
\definecolor{BlueGreen}{RGB}{49,152,255}
\definecolor{Violet}{RGB}{120,80,120}
\definecolor{Blue}{RGB}{0,0,255}
\definecolor{Yellow}{RGB}{0,255,51}
\definecolor{ElectricGreen}{RGB}{0, 255, 0}
\usepackage[unicode, colorlinks, urlcolor=BlueGreen, linkcolor=Blue, citecolor=ElectricGreen]{hyperref}

\usepackage[left=2cm,right=2cm,top=2cm,bottom=2cm,bindingoffset=0cm]{geometry}

\numberwithin{equation}{section}

\usepackage{authblk}

\title{\bf Currents of created pairs in strong electric fields}
\author[1,2]{E.~T.~Akhmedov\thanks{\href{mailto:akhmedov@itep.ru}{akhmedov@itep.ru}}}
\author[1,2]{A.~V.~Anokhin\thanks{\href{mailto:anohin.av@phystech.edu}{anohin.av@phystech.edu}}}
\author[1,2]{D.~I.~Sadekov\thanks{\href{mailto:sadekov.de@phystech.edu}{sadekov.di@phystech.edu}}}
\affil[1]{\textcolor{black}{Institutskii per, 9, Moscow Institute of Physics and Technology, 141700, Dolgoprudny, Russia}}
\affil[2]{\textcolor{black}{B. Cheremushkinskaya, 25, Institute for Theoretical and Experimental Physics, 117218, Moscow, Russia}}
\date{\today}

\begin{document}

\maketitle

\begin{abstract}

We calculate tree--level currents of created particles in strong background electric fields in 4D QED for various initial states. Namely, we do that in pulse background for initial vacuum and thermal states at past infinity. In both cases we find that the current grows linearly with the length of the pulse with coefficients of proportionality containing the characteristic Schwinger's factor. For the constant electric field background we calculate the current for several different initial states. We observe that in such a case the current is either zero or linearly divergent. We explain the reason for such a behaviour and compare the situation in ordinary and scalar QED. Finally we calculate the current in two--dimensional situation in the presence of such settings when the so called Klein paradox can be observed.

\end{abstract}

\newpage

\tableofcontents

\section{Introduction}

The proper quantities to calculate and measure in strong background fields are correlation functions rather than scattering amplitudes. Otherwise there is no cancellation of the IR divergences due to soft quanta \cite{Akhmedov:2008pu}, \cite{Akhmedov:2009vh}. Hence, tree--level scattering amplitudes are physically meaningful only when the pair production rate is very small and observed for short enough time. 

It goes without saying that in non--stationary situations (including pair creating background fields) one has to use Schwinger--Keldysh diagrammatic technique rather than the Feynman one. While the latter one is designed for the calculation of the amplitudes and corresponding cross--section, the former technique is suitable for the calculation of the correlation functions.

The goal of the present note is to calculate tree--level currents of the created pairs in various strong background electric fields in QED. The consideration of the phenomenon in question has started with the famous Klein paradox \cite{History} (see also appendix \ref{App_A}) , but we provide our considerations in a bit different settings.

Tree--level currents in QED and scalar QED has been considered in many places. For an incomplete list of references see e.g.  \cite{Gavrilov:2007hq}, \cite{Gavrilov:2007ij}, \cite{Gavrilov:2008fv}, \cite{Gavrilov:2012jk}, \cite{Gavrilov:2005dn}, \cite{Gavrilov:1996pz}, \cite{Anderson:2013ila}, \cite{Anderson:2013zia}, \cite{Mottola} and also \cite{GribBook, Schwinger}.
However, we would like to study the dependence of the behaviour of the current on the choice of the initial state. And our eventual goal is to consider loop corrections to the current. Such loop corrections in scalar QED has been considered in \cite{Akhmedov:2014hfa} and \cite{Akhmedov:2014doa}. There  observations has been made that the corrections grow with time and become comparable to the tree--level values for a long enough period of quanta creation.

We consider the four-dimensional QED in background electric fields:

\begin{equation}\label{action}
    S = \int d^4x\left(-\frac{1}{4}F_{\mu \nu}F^{\mu \nu} + \bar\psi(i\gamma^{\mu}D_{\mu} - m)\psi - j^{\mu, cl}A_{\mu}\right),
\end{equation}
We use the standard notations:
\begin{equation}
    F_{\mu \nu} = \partial_{\mu} A_{\nu} - \partial_{\nu} A_{\mu};\;\; D_{\mu} = \partial_{\mu} + i e A_{\mu}.
\end{equation}
and choose the following representation of Dirac gamma matrices:
\begin{equation}
   \gamma^{0} = \begin{pmatrix}
        1 & 0\\
        0 & -1
    \end{pmatrix};\;\;
    \gamma^{i} = \begin{pmatrix}
        o & \sigma_{i}\\
        -\sigma_{i} & 0
    \end{pmatrix}.
\end{equation}
We consider the classical background electric field, $A^{cl}_{\mu}$, which solves the Maxwell equation,
\begin{equation}
    \partial_{\mu}F^{\mu \nu, cl} = j^{\nu, cl},
\end{equation}
with an external source. In this note we consider gauge fields as classical fixed backgrounds. Only fermions are dynamical. 

We consider two types of strong electric fields and corresponding $j^{\mu, cl}$ \cite{Nikishov}. In the section two we study the pulse background, In the section three we consider the constant field background. While in the pulse background there is an obvious choice of the initial state\footnote{In the pulse background as initial states we choose the ground and thermal states for the in--modes.}, in the constant background the situation is ambiguous. In fact, in the latter situation there are no plane wave solutions. Hence, there are no positive and negative energy modes in ordinary sense. To restrict possible choices of initial states in the constant electric field background, we impose an additional constraint of proper Hadamard behaviour, as we call it here. Such a restriction physically consists in the demand that quantum fields of sufficiently large energies should be insensitive to a background. 
In appendix we use our approach to study the Klein paradox.

\section{Pulse background}\label{Pulse}

In this section we consider the electric pulse background: 

\begin{equation}
    A_{\mu}^{cl} = \left(0, 0, 0, ET\tanh\left(\frac{t}{T}\right) \right), \;\; 
    E_{\mu}^{cl} = \left(0, 0, 0, \frac{E}{\cosh^2\left(\frac{t}{T}\right)} \right).
\end{equation}
The constant electric field will be considered in the next section.

\subsection{Equations of motion}

Equations of motion for the fermion field are as follows:

\begin{equation}\label{fermeq}
    \left( i\gamma^{\mu}\partial_{\mu} - e\gamma^{\mu}A^{cl}_{\mu} - m\right)\psi(t, x) = 0.
\end{equation}
The modes \cite{Akhmedov:2009vh, Mottola, GribBook, Chervyakov:2011nr} can be represented as:

\begin{equation}\label{fourier}
    \psi_p(t, x) = \psi_{\boldsymbol{p}}(t) \, e^{i\boldsymbol{p} \boldsymbol{x}},
\end{equation}
where $\psi_{\boldsymbol{p}}(t)$ solves:

\begin{equation}\label{dirac}
    \Big[i\gamma^{0}\partial_t - (\boldsymbol{\gamma} \boldsymbol{P}) - m \Big] \, \psi_{\boldsymbol{p}}(t) = 0,
\end{equation}
and we have defined the physical momentum:

\begin{equation}
    \boldsymbol{P} = \left(p_1, \, p_2, \, p_3 + eET\tanh\left[\frac{t}{T}\right]\right).
\end{equation}
Introducing the new function $\phi_{\boldsymbol{p}}(t)$:

\begin{equation}\label{phif}
    \psi_{\boldsymbol{p}}(t) = \Big[i\gamma^{0}\partial_t - (\boldsymbol{\gamma}  \boldsymbol{P}) + m \Big] \, \phi_{\boldsymbol{p}}(t),
\end{equation}
we rewrite the equation (\ref{dirac}) as

\begin{equation}\label{equation}
    \left[ \partial^{2}_{t} + p_1^2 + p_2^2 + \left( p_3 + eET \tanh\left(\frac{t}{T}\right)  \right)^2 + m^2 + i\gamma^{0}\gamma^{3}\frac{eE}{ \cosh^2\left(\frac{t}{T} \right)} \right]\phi_{\boldsymbol{p}} (t) = 0.
\end{equation}
Finally, we look for solutions of the last equation in the form:

\begin{equation}\label{changeF}
    \phi_{p}(t) = F_{\boldsymbol{p}}^{+}(t)\chi^{1, 2};\;\; \phi_{p}(t) = F_{\boldsymbol{p}}^{-}(t)\chi^{3, 4},
\end{equation}
where

\begin{equation}\label{vectors}
    \chi^{1} = \begin{pmatrix}
        1\\
        0\\
        1\\
        0
    \end{pmatrix};\;\;
    \chi^{2} = \begin{pmatrix}
        0\\
        1\\
        0\\
        -1
    \end{pmatrix};\;\;
    \chi^{3} = \begin{pmatrix}
        0\\
        1\\
        0\\
        1
    \end{pmatrix};\;\;
    \chi^{1} = \begin{pmatrix}
        1\\
        0\\
        -1\\
        0
    \end{pmatrix},
\end{equation}
and

\begin{equation}\label{relations}
    \gamma^{0}\gamma^{3}\chi^{1, 2} = \chi^{1, 2};\;\; \gamma^{0}\gamma^{3}\chi^{3, 4} = -\chi^{3, 4}.
\end{equation}
As a result, we get that

\begin{equation}\label{modes}
    \left\{\partial^{2}_{t} + p_1^2 + p_2^2 + \left[p_3 + eET\tanh\left(\frac{t}{T} \right)\right]^2 + m^2\pm i\frac{eE}{\cosh^2\left(\frac{t}{T} \right)} \right\} \, F^{\pm}_{\boldsymbol{p}}(t) = 0.
\end{equation}
Each of the second--order differential equations (\ref{modes}) has two independent solutions, which determine the modes of the Dirac field.

\subsection{In--modes}

We choose the so-called in-modes, which correspond to asymptotically plane waves at past infinity. We will see that for them the electric current is zero at past infinity, but, then, it is generated at the future inifinity.

In the equation (\ref{modes}) it is convenient to make the change of variables $x = e^{\frac{2t}{T}}$ and introduce the following set of notations:

\begin{equation}\label{changes}
    \begin{cases}
       p_{\perp}^2 = p_1^2 + p_2^2\\
       P_{3}(t) = p_3 + eET\tanh{\left(\frac{t}{T} \right)}\\
       P_{3\pm} = p_3 \pm eET\\
       w_{+} = \sqrt{p_1^2 + p_2^2 + m^2 + (p_3 + eET)^2}\\
       w_{-} = \sqrt{p_1^2 + p_2^2 + m^2 + (p_3 - eET)^2}\\
       \theta = eET^2;\;\; \delta = 1 - i w_{-}T\\
       \beta_{-} = i\theta -\frac{i T}{2}(w_{-} + w_{+})\\
    \end{cases},
    \;\;
    \begin{cases}
    \gamma_{-} = i\theta - \frac{i T}{2}(w_{-} - w_{+})\\
       \beta_{+} = -i\theta -\frac{i T}{2}(w_{-} + w_{+})\\
       \gamma_{+} = -i\theta - \frac{i T}{2}(w_{-} - w_{+})\\
       \bar{\beta}_{-} = i\theta + \frac{i T}{2}(w_{-} - w_{+})\\
       \bar{\gamma}_{-} = i\theta + \frac{i T}{2}(w_{-} + w_{+})\\
       \bar{\beta}_{+} = -i\theta + \frac{i T}{2}(w_{-} - w_{+})\\
       \bar{\gamma}_{+} = -i\theta + \frac{i T}{2}(w_{-} + w_{+}).\\
    \end{cases}
\end{equation}
Then the solution of (\ref{modes}) can be expressed in the form:

\begin{equation}\label{pods}
    \begin{cases}
        F^{+}_{\boldsymbol{p}}(t) = e^{iw_{-}t}(1 + x)^{-i\theta}g^{+}(x)\\
        F^{-}_{\boldsymbol{p}}(t) = e^{iw_{-}t}(1 + x)^{i\theta}g^{-}(x).
    \end{cases}
\end{equation}
Where $g^\pm(x)$ obey:

\begin{equation}
    \Big\{x \, (1+x) \, \partial_x^2 + \Big[\delta + \Big(1 + \beta_{\pm} + \gamma_{\pm}\Big)\Big] \, x + \beta_{\pm}\gamma_{\pm}\Big\} \, g^{\pm}(x) = 0.
\end{equation}
Then solutions of (\ref{modes}) are:

\begin{equation*}
    F^{+}_{\boldsymbol{p}}(t) = C_1 \, e^{iw_{-}t} \, \left(1 + e^{2\frac{t}{T}}\right)^{-i\theta} \,  {}_1F_{2}\left(\beta_{+}, \gamma_{+}, \delta;-e^{2\frac{t}{T}} \right) + 
\end{equation*}
\begin{equation}\label{sol1}
     +\; C_2 \, e^{-iw_{-}t} \, \left(1 + e^{2\frac{t}{T}}\right)^{-i\theta} \,  {}_1F_{2}\left(\bar{\beta}_{+}, \bar{\gamma}_{+}, 2 - \delta;-e^{2\frac{t}{T}} \right),
\end{equation}
and

\begin{equation*}
    F^{-}_{\boldsymbol{p}}(t) = \bar{C}_1 \, e^{iw_{-}t} \, \left(1 + e^{2\frac{t}{T}}\right)^{i\theta} \,  {}_1F_{2}\left(\beta_{-}, \gamma_{-}, \delta;-e^{2\frac{t}{T}} \right) +
\end{equation*}
\begin{equation}\label{sol2}
     +\; \bar{C}_2 \, e^{-iw_{-}t}\, \left(1 + e^{2\frac{t}{T}}\right)^{i\theta} \, {}_1F_{2}\left(\bar{\beta}_{-}, \bar{\gamma}_{-}, 2 - \delta;-e^{2\frac{t}{T}} \right),
\end{equation}
where ${}_1F_{2}\left(\beta, \gamma,  \delta;z \right)$ is the hypergeometric function.

The in-modes correspond to:

\begin{equation}\label{realsol}
    \begin{cases}
        F^{+}_{\boldsymbol{p}, -}(t) = e^{-iw_{-}t} \, \left(1 + e^{2\frac{t}{T}}\right)^{-i\theta} \,  {}_1F_{2}\left(\bar{\beta}_{+}, \bar{\gamma}_{+}, 2 - \delta;-e^{2\frac{t}{T}} \right)\\
        F^{-}_{\boldsymbol{p}, +}(t) = e^{iw_{-}t} \, \left(1 + e^{2\frac{t}{T}}\right)^{i\theta} \,  {}_1F_{2}\left(\beta_{-}, \gamma_{-}, \delta;-e^{2\frac{t}{T}} \right),
    \end{cases}
\end{equation}
and behave as $F^\pm \sim e^{\mp iw_{\pm}t}$ at past infinity, i.e. as $t\to - \infty$.

Using (\ref{changeF}), (\ref{vectors}) and (\ref{phif}), for the fermion modes we get:

\begin{equation}\label{soldir1}
    \psi_{\boldsymbol{p}, 1}^{(+)}(t) = A^{+}\begin{pmatrix}
    i\partial_t - \left(P_3 - m\right)\\
    -p_1 - i p_2\\
    -i\partial_t + \left(P_3 + m\right)\\
    p_1 + i p_2
\end{pmatrix} F^{+}_{\boldsymbol{p}, -}(t), \;\; \psi_{\boldsymbol{p}, 2}^{(+)}(t) = A^{+}\begin{pmatrix}
    p_1 - i p_2\\
    i\partial_t - \left(P_3 - m\right)\\
    p_1 - i p_2\\
    i \partial_t - \left(P_3 + m\right)
\end{pmatrix} F^{+}_{\boldsymbol{p}, -}(t),
\end{equation}

\begin{equation}\label{soldir3}
    \psi_{\boldsymbol{p}, 1}^{(-)}(t) = A^{-}\begin{pmatrix}
    -p_1 + i p_2\\
    i\partial_t + \left(P_3 + m\right)\\
    p_1 - i p_2\\
    -i \partial_t - \left(P_3 - m\right)
\end{pmatrix} F^{-}_{\boldsymbol{p}, +}(t), \;\;  \psi_{\boldsymbol{p}, 2}^{(-)}(t) = A^{-}\begin{pmatrix}
    i\partial_t + \left(P_3 + m \right)\\
    p_1 + i p_2\\
    i\partial_t + \left(P_3 - m\right)\\
    p_1 + i p_2
\end{pmatrix} F^{-}_{\boldsymbol{p}, +}(t),
\end{equation}
where $P_3$ is defined in (\ref{changes}).

The normalization conditions for the obtained solutions are:

\begin{equation*}
    2 \, \bigg[\Big|\partial_t F^{+}_{\boldsymbol{p}, -}(t)\Big|^2 + \bigg(p_{\perp}^2 + P_{3}^2(t) + m^2\bigg) \, \Big|F^{+}_{\boldsymbol{p}, -}(t)\Big|^2 + 
\end{equation*}
\begin{equation}\label{norm1}
     + i P_{3}(t) \, \Big(\partial_t F^{+*}_{\boldsymbol{p}, -}(t) F^{+}_{\boldsymbol{p}, -}(t) - \partial_t F^{+}_{\boldsymbol{p}, -}(t) F^{+*}_{\boldsymbol{p}, -}(t)\Big)\bigg] = \frac{1}{|A^{+}|^2},
\end{equation}
\begin{equation*}
     2 \, \bigg[\Big|\partial_t F^{-}_{\boldsymbol{p}, +}(t)\Big|^2 + \bigg(p_{\perp}^2 + P_{3}^2(t) + m^2\bigg) \, \Big|F^{-}_{\boldsymbol{p}, +}(t)\Big|^2 -
\end{equation*}
\begin{equation}\label{norm2}
     - i P_{3}(t) \, \Big(\partial_t F^{-*}_{\boldsymbol{p}, +}(t) F^{-}_{\boldsymbol{p}, +}(t) - \partial_t F^{-}_{\boldsymbol{p}, +}(t) F^{-*}_{\boldsymbol{p}, +}(t)\Big)\bigg] = \frac{1}{|A^{-}|^2},
\end{equation}
in accordance with the canonical commutation relations, as we will see below. The normalization coefficients $A^{+}$, $A^{-}$ are time-independent, as follows from the equations of motion.

\subsection{Mode expansion of the field}

We quantize the fermion field in the standard way:

\begin{equation}\label{quant}
    \Psi_{a}(x, t) = \int \frac{d^3 p}{(2 \pi)^3}\sum_{s = 1}^{2}\left[a_{\boldsymbol{p}, s}\psi^{(+)}_{\boldsymbol{p}, s, a}(t)e^{i\boldsymbol{p}\boldsymbol{x}} + b_{\boldsymbol{p}, s}^{\dagger}\psi^{(-)}_{-\boldsymbol{p}, s, a}(t)e^{-i\boldsymbol{p}\boldsymbol{x}} \right],
\end{equation}
where:

\begin{equation}
    \left\{a_{\boldsymbol{p}, s}, {a_{\boldsymbol{k}, r}^{\dagger}} \right\} = \left\{b_{\boldsymbol{p}, s}, {b_{\boldsymbol{k}, r}^{\dagger}} \right\} = (2\pi)^3\delta(\boldsymbol{p} - \boldsymbol{k})\delta_{s, r},
\end{equation}
From these relations it follows that:

\begin{equation}
    \left\{ \Psi_{a}(x, t), \Psi_{b}^{\dagger}(y, t) \right\} = \delta(\boldsymbol{x} - \boldsymbol{y})\delta_{a,b}.
\end{equation}
In fact, anti-commutator can be represented as:

\begin{equation}\label{antic}
    \left\{ \Psi_{a}(x, t), \Psi_{b}^{\dagger}(y, t) \right\} = \int\frac{d^3 p}{(2\pi)^3}\sum_{s = 1}^{2}\left[\psi^{(+)}_{\boldsymbol{p}, s, a}(t)\psi^{(+)*}_{\boldsymbol{p}, s, b}(t) + \psi^{(-)}_{\boldsymbol{p}, s, a}(t)\psi^{(-) *}_{\boldsymbol{p}, s, b}(t) \right]e^{i\boldsymbol{p}(\boldsymbol{x} - \boldsymbol{y})}.
\end{equation}
Using the Dirac equation (\ref{dirac}) and the relation $\psi^{(-)}_{\boldsymbol{p}, s}(t) = -i\gamma_0 \gamma_2\left(\psi^{(+)}_{\boldsymbol{p}, s}(t) \right)^{*}$, one can obtain the identity:

\begin{equation}
    i\gamma^0 \partial_t \left(\sum_{s = 1}^{2}\left[\psi^{(+)}_{\boldsymbol{p}, s, a}(t)\psi^{(+)*}_{\boldsymbol{p}, s, b}(t) + \psi^{(-)}_{\boldsymbol{p}, s, a}(t)\psi^{(-) *}_{\boldsymbol{p}, s, b}(t) \right] \right) = 0.
\end{equation}
As a result, the canonical commutation relations for the fermion field follow.
The normalization constant can be found at $t = -\infty$ and are equal to:

\begin{equation}
    A \equiv |A^{\pm}| = \frac{1}{\sqrt{2p_{\perp}^2 + \left(m + P_{3-} - \sqrt{m^2 + p_{\perp}^2 + P_{3-}^2}\right)^2 + \left(m - P_{3-} + \sqrt{m^2 + p_{\perp}^2 + P_{3-}^2}\right)^2}},
\end{equation}
in accordance with the statement at the end of the previous subsection.

\subsection{Tree--level current}

Due to the transverse rotational symmetry in the electric pulse background under consideration, it is clear that only the third component of the current can be non-zero. It is equal to:

\begin{equation}\label{ordcurrent}
    \left\langle J^3 \right\rangle_{\text{tree}} \equiv \left\langle \bar{\Psi}\gamma^3 \Psi \right\rangle_{tree} = \int\frac{d^3p}{(2\pi)^3}\sum_{s = 1}^{2}\psi^{(-)\dagger}_{\boldsymbol{p}, s}(t)\gamma^{0}\gamma^{3}\psi^{(-)}_{\boldsymbol{p}, s}(t) =
\end{equation}
\begin{equation*}
     = -4|A|^2\int\frac{d^3p}{(2 \pi)^3}\bigg\{\bigg[m^2 + p_{\perp}^2 - P_3^2(t)\bigg] \, |F^{-}_{\boldsymbol{p}, +}(t)|^2 - |\partial_{t}F^{-}_{\boldsymbol{p}, +}(t)|^2 + 
\end{equation*}
\begin{equation}
    +\; iP_3(t) \, \bigg[\partial_{t}F^{-, *}_{\boldsymbol{p}, +}(t)F^{-}_{\boldsymbol{p}, +}(t) - \partial_{t}F^{-}_{\boldsymbol{p}, +}(t)F^{-, *}_{\boldsymbol{p}, +}(t)\bigg] \bigg\}.
\end{equation}
Using the second normalization condition (\ref{norm2}), one obtains that:

\begin{equation*}
     \left\langle J^3 \right\rangle_{\text{tree}} = 2\int\frac{d^3p}{(2 \pi)^3}\bigg[1 - 4|A|^2\left(m^2 + p_1^2 + p_2^2\right)|F^{-}_{\boldsymbol{p}, +}(t)|^2 \bigg] = 
\end{equation*}
\begin{equation}\label{current1}
     = 2\int\frac{d^3p}{(2 \pi)^3}\left( 1 - 4|A|^2\left(m^2 + p_{\perp}^2 \right)\left|{}_1F_{2}\left(\beta_{-}, \gamma_{-}, \delta; -e^{2\frac{t}{T}}\right)\right|^2 \right). 
\end{equation}
Expression (\ref{current1}) provides the exact tree--level current for any $t$. From it follows that the current at past infinity $t \to -\infty$ is equal to:

\begin{equation}
    \left\langle J^3 \right\rangle_{\text{tree}}^{t \to -\infty} \approx 2\int\frac{d^2p_{\perp} \, dP_{3-}}{(2 \pi)^3}\frac{-P_{3-}}{\sqrt{m^2 + p_{\perp}^2 + P_{3-}^2}} = 0,
\end{equation}
as is expected for the in--modes. 

Let us calculate now the current as $t \to +\infty$. Using the following relation for the hypergeometric function (see e.g. \cite{AkhmedovBook}):

\begin{equation*}
    {}_1F_{2}(a, b, c; x) = (-x)^{-a} \, \frac{\Gamma(c)\Gamma(b - a)}{\Gamma(c - a)\Gamma(b)} \, {}_1F_{2}\left(a, 1 + a - c, 1 + a - b; \frac{1}{x} \right) + 
\end{equation*}
\begin{equation}\label{propgeo}
    +\; (-x)^{-b} \, \frac{\Gamma(c)\Gamma(a - b)}{\Gamma(c - b)\Gamma(a)} \, {}_1F_{2}\left(b, 1 + b - c, 1 + b - a; \frac{1}{x} \right),\quad a - b \notin \mathds{Z},
\end{equation}
one finds that:

\begin{equation}
    {}_1F_{2}\Big(a, b, c; x \rightarrow -\infty\Big) \approx (-x)^{-a} \, \frac{\Gamma(c)\Gamma(b - a)}{\Gamma(c - a)\Gamma(b)} + (-x)^{-b} \, \frac{\Gamma(c)\Gamma(a - b)}{\Gamma(c - b)\Gamma(a)}.
\end{equation}
It allows one to simplify the expression for the current at $t \to +\infty$:

\begin{equation*}
    \left\langle J^3 \right\rangle_{\text{tree}}^{t \to +\infty} \approx 2\int \frac{p_{\perp}dp_{\perp} dp_3}{(2\pi)^2}\Bigg\{1 - (m^2 + p_{\perp}^2)\times
\end{equation*}
\begin{equation}\label{current01}
    \times\frac{w_{-} \left[-w_{+} + P_{3+}\left(\coth{(\pi T w_{-})}\coth{(\pi T w_{+})} - \frac{\cosh{(2\pi EeT^2)}}{\sinh{(\pi T w_{-})}\sinh{(\pi T w_{+})}}\right)\right]}{w_{+}(P_{3-} - w_{-}) \, \left[m^2 + p_{\perp}^2 + P_{3-}(P_{3-} - w_{-})\right]}\Bigg\}.
\end{equation}
The integral (\ref{current01}) is convergent, and the main contribution to the $p_3$ integral is coming from $|p_3| < eET$. We estimate the current in the approximation of long pulse $T$ and small mass $m$:

\begin{equation}\label{limits}
    eET^2 \gg 1, \;\; eET \gg m.
\end{equation}
Changing the variables $p_3 \rightarrow P_{3-}$ in (\ref{current01}) and using
\begin{equation}\label{proptanh}
    \lim_{x \rightarrow \infty} \frac{\cosh{(x + a + b)}\cosh{(x + a - b)} - \cosh{(2x)}}{\sinh{(x + a + b)}\sinh{(x + a - b)}}  - 1 = -2e^{-2a},
\end{equation}
one can find that:

\begin{equation}\label{res02}
    \left\langle J^3 \right\rangle_{\text{tree}}^{t \to +\infty} \simeq \frac{2}{(2 \pi)^2}\int_{0}^{+\infty}dp_{\perp} \int_{-eET}^{eET}dp_3 \left(e^{-\frac{\pi m^2}{eE}}e^{-\frac{\pi p_{\perp}^2}{eE}}p_{\perp}\right) \simeq \frac{(eE)^2T}{2 \pi^3}e^{-\frac{\pi m^2}{eE}}
\end{equation}
up to terms of the highest order in $\frac{1}{T}$. This result agrees with the one found in \cite{Gavrilov:2007hq}, \cite{Gavrilov:2007ij}, \cite{Gavrilov:2008fv}, \cite{Gavrilov:2012jk}, \cite{Gavrilov:2005dn}, \cite{Gavrilov:1996pz}, \cite{Anderson:2013ila}, \cite{Anderson:2013zia}, \cite{Mottola}, \cite{Zahn:2015awa} in ordinary and scalar QED.

\subsection{Tree--level current for the thermal initial state}\label{ThermalCurrent}

In this section we consider the thermal initial state at past infinity:

\begin{equation}\label{tempa1}
     \langle a^{\dagger}_{\boldsymbol{p}, s} a_{\boldsymbol{k}, r}  \rangle = \langle b^{\dagger}_{\boldsymbol{p}, s} b_{\boldsymbol{k}, r} \rangle = (2\pi)^3 \, \frac{\delta_{s, r} \, \delta(\boldsymbol{p} - \boldsymbol{k})}{e^{\beta w(\boldsymbol{p}, t)} + 1} \equiv (2\pi)^3 \, \delta_{s, r} \, \delta(\boldsymbol{p} - \boldsymbol{k}) \,  n(\boldsymbol{p}, t),
\end{equation}
\begin{equation}\label{tempa2}
     \langle a_{\boldsymbol{p}, s} a^{\dagger}_{\boldsymbol{k}, r}  \rangle =  \langle b_{\boldsymbol{p}, s} b^{\dagger}_{\boldsymbol{k}, r}  \rangle = (2\pi)^3 \, \frac{\delta_{s, r} \, \delta(\boldsymbol{p} - \boldsymbol{k}) \, e^{\beta w(\boldsymbol{p}, t)}}{e^{\beta w(\boldsymbol{p}, t)} + 1} \equiv (2\pi)^3 \, \delta_{s, r} \, \delta(\boldsymbol{p} - \boldsymbol{k}) \, \Big[1 - n(\boldsymbol{p}, t)\Big].
\end{equation}
Here $\beta$ is the inverse temperature and 
$$
w(\boldsymbol{p}, t) = \sqrt{p_1^2 + p_2^2 + \left[p_3 + eET\tanh\left(\frac{t}{T} \right)\right]^2 + m^2}
$$
is the dispersion relation in the background field.
With such a modification the expression for the tree--level current (\ref{ordcurrent}) is:

\begin{equation*}
     \left\langle J^3 \right\rangle_{\text{tree}, \; T} = 2\int\frac{d^3p}{(2 \pi)^3} \, \Big[1 - n(\boldsymbol{p}, t)\Big] \, \bigg[1 - 4|A|^2\left(m^2 + p_{\perp}^2\right)|F^{-}_{\boldsymbol{p}, +}(t)|^2 \bigg] -
\end{equation*}
\begin{equation*}
    -\; 2\int\frac{d^3p}{(2 \pi)^3} \, n(\boldsymbol{p}, t) \, \bigg[1 - 4|A|^2\left(m^2 + p_{\perp}^2\right)|F^{+}_{\boldsymbol{p}, -}(t)|^2 \bigg] = 
\end{equation*}
\begin{equation}\label{termcurrent1}
     = 2\int\frac{d^3p}{(2 \pi)^3} \, \Big[1 - 2n(\boldsymbol{p}, t)\Big] \, \left[ 1 - 4|A|^2\left(m^2 + p_{\perp}^2 \right)\left|{}_1F_{2}\left(\beta_{-}, \gamma_{-}, \delta, -e^{2\frac{t}{T}}\right)\right|^2 \right].
\end{equation}
We are interested in the value of the current at past, $t \to -\infty$, and future, $t \to +\infty$, infinities. In these limits, the dispersion relation reduces to:

\begin{equation}\label{spectrum}
    w^2(\boldsymbol{p}, t \to \pm \infty) \approx w_\pm^2 = p_1^2 + p_2^2 + P_{3\pm}^2 + m^2,
\end{equation}
where $w_\pm$ and $P_{3\pm}$ are defined in (\ref{changes}).
Hence:

\begin{equation}\label{factor}
    1 - 2n(\boldsymbol{p}, t \to \pm \infty) \approx \tanh\left(\frac{\beta \sqrt{p_{\perp}^2 + m^2 + P_{3\pm}^2}}{2} \right).
\end{equation}
Then the expression for the current at $t \to -\infty$ is as follows:

\begin{equation}
    \left\langle J^3 \right\rangle_{\text{tree}, \;T}^{t \to -\infty} \approx 2\int\frac{d^2p_\perp \, dP_{3-}}{(2 \pi)^3}\frac{-P_{3-}\tanh\left(\frac{\beta \sqrt{p_{\perp}^2 + m^2 + P_{3-}^2}}{2} \right)}{\sqrt{m^2 + p_{\perp}^2 + P_{3-}^2}} = 0.
\end{equation}
At the same time at $t \to + \infty$ it has the form:

\begin{equation*}
    \left\langle J^3 \right\rangle_{\text{tree}, \;T}^{t \to +\infty} \approx 2\int \frac{p_{\perp}dp_{\perp} dp_3}{(2\pi)^2}\tanh\left(\frac{\beta \sqrt{p_{\perp}^2 + m^2 + P_{3+}^2}}{2} \right)\Bigg\{1 - (m^2 + p_{\perp}^2)\times
\end{equation*}
\begin{equation}\label{termcurrent01}
    \times\frac{w_{-} \, \left[-w_{+} + P_{3+}\left(\coth{(\pi T w_{-})}\coth{(\pi T w_{+})} - \cosh{(2\pi EeT^2)}\frac{1}{\sinh{(\pi T w_{-})}\sinh{(\pi T w_{+})}}\right)\right]}{w_{+} \, (P_{3-} - w_{-}) \, \Big[m^2 + p_{\perp}^2 + P_{3-}(P_{3-} - w_{-})\Big]}\Bigg\}.
\end{equation}
Let us investigate the low and high temperature limiting expressions for the current. In the limit (\ref{limits}) and at low temperatures $\beta eET \gg 1$, we obtain:
\begin{equation*}
    \left\langle J^3 \right\rangle_{\text{tree}, \;T}^{t \to +\infty} \simeq \frac{2}{(2 \pi)^2}\int_{0}^{+\infty}dp_{\perp} \int_{-eET}^{eET}dp_3 \left(e^{-\frac{\pi m^2}{eE}}e^{-\frac{\pi p_{\perp}^2}{eE}}p_{\perp}\right)\left(1 - 2 e^{-\frac{\beta eET}{2}\left(1 + \frac{p_{\perp}^2}{2(eET)^2}\right)}\right) \simeq
\end{equation*}
\begin{equation}\label{termres02big}
    \simeq \frac{(eE)^2T}{2 \pi^3}e^{-\frac{\pi m^2}{eE}}\left(1 - 2 \frac{\exp{\left(-\frac{\beta eET}{2}\right)}}{1 + \frac{\beta}{4 \pi T}} \right),
\end{equation}
which reduces to the previously obtained result, if we put $\beta$ to infinity.

At the same time, in the case of high temperatures $\beta m \ll 1$, $\beta eET \ll 1$ the current is:

\begin{equation*}
    \left\langle J^3 \right\rangle_{\text{tree}, \;T}^{t = +\infty} \simeq \frac{2}{(2 \pi)^2}\int_{0}^{+\infty}dp_{\perp} \int_{-eET}^{eET}dp_3 \left(e^{-\frac{\pi m^2}{eE}}e^{-\frac{\pi p_{\perp}^2}{eE}}p_{\perp}\right)\frac{\beta p_{\perp}}{2} \simeq
\end{equation*}
\begin{equation}\label{termres02small}
    \simeq \frac{(eE)^{\frac{5}{2}}\beta T}{8 \pi^3}e^{-\frac{\pi m^2}{eE}},
\end{equation}
i.e. the rate of the growth is changed a bit, but still contains the characteristic exponential factor of Schwinger's type. The current with the finite temperature initial states was estimated from the Bogoliubov transformation in \cite{Kim:2008em}.

\section{Constant electric field background}\label{ConstantEF}

In this section we consider the QED in the constant electric field background in (3+1) dimensions. We choose the temporal gauge $A_3 = eEt$ and use the same representation of gamma matrices as in the previous section. 

\subsection{Modes}\label{SModes}

In the constant electric field background one obtains:
\begin{equation}\label{F6}
	\left( \partial_{t}^{2} + P_{3}^{2} + \left| p_{\perp} \right| ^2 + m^2 + ieE \gamma^{0}\gamma^{3}   \right) \phi_{\boldsymbol{p}}(t) = 0,
\end{equation}
in the same way as eq. (\ref{equation}) was obtained. Here we denote $p_{\perp} = p_{1} + ip_{2}$ and $P_3 = p_3+eEt$. In the case under consideration the modes are as follows \cite{Mottola}:
\begin{equation}\label{F9}
\begin{array}{ll}
\psi_{\textbf{p},1}^{(+)}(t) =
\begin{pmatrix}
i\partial_t f_{\textbf{p}}(t) + (m-P_3)f_{\textbf{p}}(t)\\
-p_{\perp} f_{\textbf{p}}(t)\\
-i\partial_t f_{\textbf{p}}(t) + (m+P_3)f_{\textbf{p}}(t)\\
p_{\perp} f_{\textbf{p}}(t)
\end{pmatrix}, &
	\psi_{\textbf{p},2}^{(+)}(t) = 
\begin{pmatrix}
p_{\perp}^* f_{\textbf{p}}(t)\\
i\partial_t f_{\textbf{p}}(t) + (m-P_3)f_{\textbf{p}}(t)\\
p_{\perp}^* f_{\textbf{p}}(t)\\
i\partial_t f_{\textbf{p}}(t) - (m+P_3)f_{\textbf{p}}(t)
\end{pmatrix},
\end{array}
\end{equation}
\begin{equation}\label{F10}
\begin{array}{ll}
	\psi_{\textbf{p},1}^{(-)}(t) = 
\begin{pmatrix}
-p_{\perp}^* \varphi_{\textbf{p}}(t)\\
i\partial_t \varphi_{\textbf{p}}(t) + (m+P_3)\varphi_{\textbf{p}}(t)\\
p_{\perp}^* \varphi_{\textbf{p}}(t)\\
-i\partial_t \varphi_{\textbf{p}}(t) + (m-P_3)\varphi_{\textbf{p}}(t)
\end{pmatrix},  &
\psi_{\textbf{p},2}^{(-)}(t) = 
\begin{pmatrix}
i\partial_t \varphi_{\textbf{p}}(t) + (m+P_3)\varphi_{\textbf{p}}(t)\\
p_{\perp} \varphi_{\textbf{p}}(t)\\
i\partial_t \varphi_{\textbf{p}}(t) - (m-P_3)\varphi_{\textbf{p}}(t)\\
p_{\perp} \varphi_{\textbf{p}}(t)
\end{pmatrix},
\end{array}
\end{equation}
where the functions $f_{\boldsymbol{p}}(t), \varphi_{\boldsymbol{p}}(t)$ satisfy the equations as follows

\begin{equation}\label{F11}
\begin{array}{ll}
	\left[\partial_{z}^2 + \left(\nu+\frac{1}{2}-\frac{z^2}{4}\right)\right]f_{\textbf{p}}(t) = 0, &
	\left[\partial_{z}^2 + \left(\nu-\frac{1}{2}-\frac{z^2}{4}\right)\right]\varphi_{\textbf{p}}(t) = 0,
\end{array} 
\end{equation}
$$\nu=-i\frac{m^2+|p_{\perp}|^2}{2eE}, \;\; z = e^{i\frac{\pi}{4}}\sqrt{2eE} \, \left(t+\frac{p_3}{eE}\right) = e^{i\frac{\pi}{4}}\sqrt{\frac{2}{eE}}P_3.$$
Solutions of the equations (\ref{F11}) are  Weber's parabolic cylinder functions \cite{Gavrilov:2007hq, Gavrilov:2015zem}. Hence, the simplest choice of modes is

\begin{equation}\label{F12}
	        \begin{cases}
			f_{\mathbf{p}}(t)= \alpha D_{\nu}(z) + \beta D_{\nu}(-z),\\
			\varphi_{\textbf{p}}(t)= \tilde{\alpha} D_{-\nu}(-iz) + \tilde{\beta} D_{-\nu}(iz).
			\end{cases}
\end{equation}
Please note that: $f_{\textbf{p}}(t) = f_{p_{\perp}}(P_3), \; \varphi_{\textbf{p}}(t) = \varphi_{p_{\perp}}(P_3)$ and $\psi_{\textbf{p}s}^{ \left( \pm \right)} \left( t \right) \equiv \psi_{p_{\perp}s}^{\left( \pm \right)} \left( P_{3} \right)$, where $P_3$ is defined in the second line of (\ref{F11}). Note that $z\sim P_3$ and in the constant electric field background there is the time--translation invariance under the simultaneous shifts $t\to t+a$ and $p_3 \to p_3 - eEa$. Also note that the relation $\psi_{\textbf{p}s}^{\left( - \right)} \left( t \right) = - i \gamma^0 \gamma^{2} (\psi_{\textbf{p}s}^{\left( + \right)}  \left( t \right) )^{*}$, which is valid even in the vanishing background electric field case as well.

To restrict the choice of the modes and states in the constant electric field background, we impose the requirement of the proper Hadamard behaviour. I.e. we demand that the modes turn into the ordinary plane waves in the limit of high momentum $|\boldsymbol{p}| \rightarrow \infty$ for constant $t$:
\begin{equation}\label{adam}
    \psi_{\boldsymbol{p}}(t) \sim e^{-i|\boldsymbol{p}|t}e^{i\boldsymbol{p} \boldsymbol{x}}.
\end{equation}
Such a restriction is required for the correct behavior of the propagators in the UV limit, which is necessary for the independence of the leading order ultraviolet renormalizations from the value of the background field.

In this section we investigate the dependence of the tree--level current on the coefficients $\alpha, \beta, \tilde{\alpha}$ and $\tilde{\beta}$ in (\ref{F12}). One can use the asymptotics of the parabolic cylinder functions \cite{Akhmedov:2019rvx, Crothers, Bateman} to find the coefficients $\alpha, \beta, \tilde{\alpha}$ and $\tilde{\beta}$ corresponding to the proper Hadamard behaviour and get the functions as follows:

\begin{equation}\label{F29}
	f_{\textbf{p}}^{(+)}(t) = A_{\textbf{p}}^{(+)}\left(D_{\nu}(z)-e^{-\pi \frac{m^2+|p_{\perp}|^2}{2eE}}D_{\nu}(-z)\right),
\end{equation}
\begin{equation}\label{F30}
	\varphi_{\textbf{p}}^{(-)}(t) = A_{\textbf{p}}^{(-)}\left(D_{-\nu}(-iz)-e^{-\pi\frac{m^2+|p_{\perp}|^2}{2eE}}D_{-\nu}(iz)\right) = A_{\textbf{p}}^{(-)}\left(D_{\nu}^*(z)-e^{-\pi\frac{m^2+|p_{\perp}|^2}{2eE}}D_{\nu}^*(-z)\right),
\end{equation}
which correspond to the positive and negative frequency solutions in the UV limit $|\boldsymbol{p}|\rightarrow \infty$ at constant $t$.

\subsection{Mode expansion of the field}\label{SSFer_field_expansion}

We decompose the fermion field in the same way as in (\ref{quant}), using the modes defined in the previous subsection (\ref{F12}), (\ref{F9})-(\ref{F10}).
We need to find such coefficients $\alpha, \beta, \tilde{\alpha}$ and $\tilde{\beta}$ in (\ref{F12}) that both

\begin{equation}\label{F32}
	\left\{ a_{\textbf{p}, s}, a_{\textbf{k}, r}^{\dagger}  \right\} = \left\{ b_{\textbf{p}, s}, b_{\textbf{k}, r}^{\dagger}  \right\} = (2\pi)^3\delta(\textbf{p} - \textbf{k} ) \delta_{s}^{r} ,
\end{equation}
and

$$
\left\{ \Psi(t, \textbf{x})_{a} ,  \Psi(t, \textbf{y})_{b}^{\dagger}  \right\} =
	 	\int \frac{d^{2} p_{\perp}}{\left(2 \pi \right)^{2}} \int \frac{dP_3}{2 \pi} \sum_{s=1}^{2} \left[ \psi_{p_{\perp}s,a}^{\left( + \right)} \left( P_{3} \right)  \psi_{p_{\perp}s,b}^{\left( + \right)*}\left( P_{3} \right) + \psi_{p_{\perp}s,a}^{\left( - \right)}\left( P_{3} \right) \psi_{p_{\perp}s,b}^{\left( - \right)*}\left( P_{3} \right) \right] e^{i\textbf{p}(\textbf{x} - \textbf{y})} =
$$
\begin{equation}\label{F33}
	= \delta(\textbf{x} - \textbf{y} ) \delta_{a}^{b}, 
\end{equation}
canonical commutation relations are satisfied.

To fulfil the last requirement the coefficients in (\ref{F12}) should satisfy the following constraints

\begin{equation}\label{F35}
\begin{cases}
\alpha \beta^* - \tilde{\alpha}^* \tilde{\beta}=0,\\
|\alpha|^2=|\tilde{\alpha}|^2,\;|\beta|^2=|\tilde{\beta}|^2,\\
|\alpha|^2 + |\beta|^2 +2\text{Re}(\alpha^*\beta)e^{-\pi\frac{m^2+|p_{\perp}|^2}{2eE}} = \frac{e^{-\pi \frac{m^{2} +  \left|p_{\perp} \right|^{2}}{4eE}}}{2(m^2+|p_{\perp}|^2)}.
\end{cases}
\end{equation}
Due to the first two conditions of (\ref{F35}) one can see that $|\varphi_{\boldsymbol{p}}(t)| = |f_{\boldsymbol{p}}(t)|$. We will use this relation below.

We emphasize here that the all relations (\ref{F32})-(\ref{F35}) are time independent.
For the mode functions (\ref{F29})-(\ref{F30}) these conditions imply that

\begin{equation}\label{F36}	
	|A_{\textbf{p}}^{(+)}|^2 = 	|A_{\textbf{p}}^{(-)}|^2 = \frac{1}{2(m^2+|p_{\perp}|^2)}\frac{e^{-\pi \frac{m^{2} +  \left|p_{\perp} \right|^{2}}{4eE}}}{1-e^{-\pi \frac{m^{2} +  \left|p_{\perp} \right|^{2}}{eE}}}.
\end{equation}
Also it is instructive to rewrite the normalization conditions via the  functions $f_{\textbf{p}}(t)$ and $\varphi_{\textbf{p}}(t)$ as:

\begin{equation}\label{F37}
	\Big|\partial_{t}f_{\textbf{p}}\Big|^2 + \bigg(m^2+|p_{\perp}|^2+P_3^2\bigg) \, \Big|f_{\textbf{p}}\Big|^2 + iP_3 \, \bigg(f_{\textbf{p}}\partial_{t}f_{\textbf{p}}^* - \partial_{t}f_{\textbf{p}}f_{\textbf{p}}^*\bigg) = \frac12,
\end{equation}
\begin{equation}\label{F38}
\Big|\partial_{t}\varphi_{\textbf{p}}\Big|^2 + \bigg(m^2+|p_{\perp}|^2+P_3^2\bigg) \, \Big|\varphi_{\textbf{p}}\Big|^2 - iP_3 \, \bigg(\varphi_{\textbf{p}}\partial_{t}\varphi_{\textbf{p}}^* - \partial_{t}\varphi_{\textbf{p}}\varphi_{\textbf{p}}^*\bigg) = \frac12.
\end{equation}
We will use these relations below.

\subsection{Tree-level current}\label{STree_current}

The tree--level current can be found in the same way as it was done in the previous section:

\begin{gather}
	\left< J^{3} \right>_{\text{tree}} \equiv \left< \overline{\Psi} \gamma^{3} \Psi \right> = 
	-4 \int \frac{d^3 \textbf{p}}{(2\pi)^3} \bigg[ \Big(m^2 + |p_{\perp}|^2 - P_3^2\Big) \, \Big|f_{\textbf{p}}\Big|^2 - \Big|\partial_t f_{\textbf{p}}\Big|^2 - iP_3 \, \Big(f_{\textbf{p}} \partial_t f^*_{\textbf{p}} - \partial_t f_{\textbf{p}} f^*_{\textbf{p}} \Big) \bigg] =
		 \nonumber \\ =
			2 \int \frac{d^3 \textbf{p}}{(2\pi)^3} \bigg[ 1 - 4 \Big(m^2+|p_{\perp}|^2 \Big) \, \Big|f_{\textbf{p}}\Big|^2 \bigg],
\label{F39}
\end{gather}
where $f_{\textbf{p}}$ is defined in (\ref{F12}). To obtain this relation we have used that $|\varphi_{\boldsymbol{p}}(t)| = |f_{\boldsymbol{p}}(t)|$, as was explained in the previous subsection.

With the use of equations of motion and normalization conditions (\ref{F37})-(\ref{F38}) one can prove the following expression for the current:

\begin{equation}\label{CurrentExpression}
\langle J^{3} \rangle_{\text{tree}} = \int \frac{d^{2} p_{\perp}}{\left(2 \pi \right)^{2}} \int^\Lambda_{-\Lambda} \frac{dP_3}{2 \pi} \, \partial_{P_3} {\mathcal{F}_{p_{\perp}}(P_3)}, 
\end{equation}
where

\begin{equation}
\mathcal{F}_{p_{\perp}}(P_3) = 2 \left\{ \frac{m^2+|p_{\perp}|^2 + P_3^2}{P_3} - \frac{2 \, (m^2 + |p_{\perp}|^2)}{P_3} \bigg[ \Big(m^2+|p_{\perp}|^2 + P_3^2\Big) \, \Big|f\Big|^2 + \Big|\partial_t f\Big|^2 \bigg] \right\},
\end{equation}
and $\Lambda$ is the cutoff of the physical momentum.

We stress that (\ref{CurrentExpression}) consistently relies on the time--translation invariance, which is present in the constant electric field background and reveals itself in the fact that $f_{\textbf{p}}(t) = f_{p_{\perp}}(P_3)$. This allows one to write

\begin{equation}\label{F41}
\left<J^3\right>_{\text{tree}} = 2\underset{\Lambda \rightarrow +\infty}{\lim}\int \frac{d^{2} p_{\perp}}{\left(2 \pi \right)^{3}} \left\{ \frac{m^2+|p_{\perp}|^2 + P_3^2}{P_3} - \frac{2(m^2 + |p_{\perp}|^2)}{P_3} \, \bigg[ \Big(m^2+|p_{\perp}|^2 + P_3^2\Big) \, \Big|f\Big|^2 + \Big|\partial_t f\Big|^2  \bigg] \right\}\bigg|_{-\Lambda}^{\Lambda}.
\end{equation}
Using again the asymptotics of the parabolic cylinder functions \cite{Crothers, Bateman} and the differentiation formulas, we find that the integrand in (\ref{F41}) behaves as:
\begin{gather}
	\left\{ \frac{m^2+|p_{\perp}|^2 + P_3^2}{P_3} - \frac{2(m^2 + |p_{\perp}|^2)}{P_3} \, \bigg[ \Big(m^2+|p_{\perp}|^2 + P_3^2\Big) \, \Big|f\Big|^2 + \Big|\partial_t f\Big|^2  \bigg] \right\}\bigg|_{-\Lambda}^{\Lambda}= \nonumber \\
	=-2\Lambda e^{-\pi\frac{m^2+|p_{\perp}|^2}{eE}} \frac{|\alpha|^2 + |\beta|^2 +2\text{Re}(\alpha^*\beta)e^{\pi\frac{m^2+|p_{\perp}|^2}{2eE}}}{|\alpha|^2 + |\beta|^2 +2\text{Re}(\alpha^*\beta)e^{-\pi\frac{m^2+|p_{\perp}|^2}{2eE}}} + \ldots,
\label{F48}
\end{gather}
where ellipsis stand for terms, which are vanishing in the limit $\Lambda \rightarrow +\infty$. Note that there is no finite contribution to the last expression. We will explain the reason for that below.

As a result, we find:

\begin{gather}
\left<J^3\right>_{\text{tree}} = -4\underset{\Lambda \rightarrow +\infty}{\lim} \Lambda \int \frac{d^{2} p_{\perp}}{\left(2 \pi \right)^{3}} \left[e^{-\pi\frac{m^2+|p_{\perp}|^2}{eE}} \frac{|\alpha|^2 + |\beta|^2 +2\text{Re}(\alpha^*\beta)e^{\pi\frac{m^2+|p_{\perp}|^2}{2eE}}}{|\alpha|^2 + |\beta|^2 +2\text{Re}(\alpha^*\beta)e^{-\pi\frac{m^2+|p_{\perp}|^2}{2eE}}} \right] = \nonumber \\
=-8 \underset{\Lambda \rightarrow +\infty}{\lim} \Lambda \int \frac{d^{2} p_{\perp}}{\left(2 \pi \right)^{3}} \left[e^{-3\pi\frac{m^2+|p_{\perp}|^2}{4eE}}(m^2+|p_{\perp}|^2)\left(|\alpha|^2 + |\beta|^2 +2\text{Re}(\alpha^*\beta)e^{\pi\frac{m^2+|p_{\perp}|^2}{2eE}}\right) \right].
\label{F49}
\end{gather}
For the modes with the proper Hadamard behaviour (\ref{F29}), (\ref{F30}) the current is divergent

\begin{equation}\label{F50}
	\langle J^3 \rangle_{\text{tree}} = \underset{\Lambda \rightarrow +\infty}{\lim}\frac{e^{-\pi\frac{m^2}{eE}}}{2\pi^3}eE\Lambda.
\end{equation}
Compare this result with the one for $f_{\textbf{p}}(t) = A D_{\nu}(z)$, i.e. when $\beta = 0$ in (\ref{F12}):

\begin{equation}\label{F51}
\langle J^3 \rangle_{\text{tree}} = -\underset{\Lambda \rightarrow +\infty}{\lim}\frac{e^{-\pi\frac{m^2}{eE}}}{2\pi^3}eE\Lambda.
\end{equation}
The current (\ref{F49}) is linearly divergent, which agrees with the result in the pulse background. The latter one is growing with the length of the pulse and becomes infinite for the constant field --- infinitely long pulse. 

There is also a possibility that the current (\ref{F49}) is vanishing, when the following condition holds:

\begin{equation}\label{F52}
	|\alpha|^2 + |\beta|^2 +2\text{Re}(\alpha^*\beta)e^{\pi\frac{m^2+|p_{\perp}|^2}{2eE}} = 0.
\end{equation}
Using the parametrization $\alpha = A \cos(\varphi), \; \beta = A e^{i\theta} \sin(\varphi)$ in (\ref{F35}) and (\ref{F52}) the mode functions, which give the vanishing current, can be written as follows:

\begin{gather}\label{F53}
	f_{\textbf{p}}(t)=A\bigg(D_{\nu}(z)\cos(\varphi) - D_{\nu}(-z)\sin(\varphi)\bigg), \\
	|A|^2 = \frac{1}{2(m^2+|p_{\perp}|^2)} \frac{e^{-\pi\frac{m^2+|p_{\perp}|^2}{4eE}}}{1-e^{-\pi \frac{m^2+|p_{\perp}|^2}{eE}}},\; \varphi = \frac{1}{2} \arcsin \left(e^{-\pi \frac{m^2+|p_{\perp}|^2}{2eE}}\right). \nonumber
\end{gather}
These modes violate the condition of proper Hadamard  behaviour. However, we consider them to discuss the situation when the current does vanish. In the literature these modes are frequently discussed without stressing the fact that they violate the condition of proper Hadamard  behaviour.

To understand the situation better we compare the situation in QED with fermions to the current in the constant electric field background in the scalar QED.

\section{Comment about the theory with scalars instead of fermions}\label{Bosons}

We want to compare our observations with the situation in the scalar QED.
Namely, consider massive scalar field theory \cite{Anderson:2013ila, Anderson:2013zia, Krotov:2010ma}
\begin{equation}\label{F54}
	S = \int d^4x \left[ -\frac{1}{4}F_{\mu\nu}^2 + |D_{\mu}\phi|^2 - m^2 \phi^2 - j_{\mu}^{cl}A^{\mu} \right]
\end{equation}
with the temporal gauge for the classical background \cite{Akhmedov:2014hfa, Akhmedov:2014doa}. The scalar field can  be decomposed as
\begin{equation}\label{F55}
	\phi(\textbf{x},t) = \int \frac{d^3\textbf{p}}{(2\pi)^2} \left[ a_{\textbf{p}}f_{\textbf{p}}(t)e^{i\textbf{px}} + b^{\dagger}_{\textbf{p}}f_{-\textbf{p}}^*(t)e^{-i\textbf{px}} \right],
\end{equation}
where the mode functions obey the Klein--Gordon equation, and

\begin{equation}\label{F56}
	\left[\frac{d^2}{dt^2}+\omega_{\textbf{p}}^2(t)\right] f_{\textbf{p}}(t) = 0, \quad {\rm where} \quad \omega_{\textbf{p}}^2 = m^2+|p_{\perp}|^2+P_3^2.
\end{equation}
The solution to the last equation (in the same notations of the previous subsection) is

\begin{equation}\label{F57}
	f_{\textbf{p}}(t) = \alpha D_{\nu-\frac{1}{2}}(z) + \beta D_{\nu-\frac{1}{2}}(-z).
\end{equation}
Normalization condition following from the commutation relations $\left[\phi(\textbf{x},t), \partial_{t} \phi^*(\textbf{y},t) \right] = i\delta(\textbf{x}-\textbf{y})$ imposes the constraint
\begin{equation}\label{F58}
	|\alpha|^2 - |\beta|^2 - 2\text{Im}(\alpha^*\beta)e^{-\pi\frac{m^2+|p_{\perp}|^2}{2eE}} = \frac{e^{-\pi\frac{m^2 + |p_{\perp}|^2}{4eE}}}{\sqrt{2eE}}.
\end{equation}
The current in this theory,
\begin{equation}\label{F59}
	J_{\mu}^{\text{\tiny scalar}} = i\left( \phi^*D_{\mu}^{cl} \phi - \phi D_{\mu}^{cl}\phi^* \right),
\end{equation}
at tree--level is equal to:

\begin{equation}\label{F60}
	\langle J^3_{\text{\tiny scalar}}\rangle_{\text{tree}} = \int \frac{d^3\textbf{p}}{(2\pi)^3} \, P_3 \, \big|f_{\textbf{p}}(t)\big|^2 = \int \frac{d^2p_{\perp}}{2(2\pi)^3}\int dP_3 \, \partial_{P_3} \bigg[ \Big(m^2+|p_{\perp}|^2+P_3^2\Big) \, \Big|f_{\textbf{p}}(t)\Big|^2 + \Big|\partial_t f_{\textbf{p}}(t)\Big|^2 \bigg].
\end{equation}
In the same way as for fermions we get that

\begin{equation}\label{F61}
	\langle J^3_{\text{\tiny scalar}}\rangle_{\text{tree}} = -\underset{\Lambda \rightarrow + \infty}{\lim}\sqrt{2eE}\Lambda \int \frac{d^2p_{\perp}}{(2\pi)^3} \left[e^{-3\pi\frac{m^2+|p_{\perp}|^2}{4eE}}\left(|\alpha|^2 - |\beta|^2 +2\text{Im}(\alpha^*\beta)e^{\pi\frac{m^2+|p_{\perp}|^2}{2eE}}\right) \right].
\end{equation}
Again we see, that the current is either zero or is linearly divergent as $\Lambda \to \infty$. In this case the condition when the current vanishes is
\begin{equation}\label{F62}
	|\alpha|^2 - |\beta|^2 +2\text{Im}(\alpha^*\beta)e^{\pi\frac{m^2+|p_{\perp}|^2}{2eE}} = 0.
\end{equation}
Compare this to eq. (\ref{F52}) for fermions. This condition can be solved as e.g. $\alpha=1, \; \beta = 1$ or $\alpha=1, \; \beta = -1$. Such choices correspond to the so called fundamental real solutions of (\ref{F56}) \cite{Anderson:2013zia, Bateman}:
\begin{equation}\label{F63}
	f_{p_{\perp}}^{(0)} = 2^{-\frac{\nu}{2} - \frac{3}{4}} \frac{\Gamma\left( -\frac{\nu}{2} + \frac{3}{4} \right)}{\sqrt{\pi}} \left[ D_{\nu-\frac{1}{2}}(z) + D_{\nu-\frac{1}{2}}(-z) \right],
\end{equation}
\begin{equation}\label{F64}
f_{p_{\perp}}^{(1)} = 2^{-\frac{\nu}{2} - \frac{5}{4}} e^{-i\frac{\pi}{4}} \frac{\Gamma\left( -\frac{\nu}{2} + \frac{1}{4} \right)}{\sqrt{\pi}} \left[ -D_{\nu-\frac{1}{2}}(z) + D_{\nu-\frac{1}{2}}(-z) \right],
\end{equation}
where $z$ is defined in (\ref{F11}).
However, these functions do not obey the condition (\ref{F58}). But one can solve (\ref{F62}) also by the following expressions:

\begin{equation}\label{F65}
	\alpha = A\left( 1 - k e^{\pi\frac{m^2+|p_{\perp}|^2}{4eE}} + ik e^{-\pi\frac{m^2+|p_{\perp}|^2}{4eE}} \right),
\end{equation}
\begin{equation}\label{F66}
\beta = A\left( 1 + k e^{\pi\frac{m^2+|p_{\perp}|^2}{4eE}} - ik e^{-\pi\frac{m^2+|p_{\perp}|^2}{4eE}} \right),
\end{equation}
where $k \in \mathbb{R}$. In such a case, this solution obeys both (\ref{F58}) and (\ref{F62}) and has the property $v_{p_{\perp}}(-P_3) = v_{p_{\perp}}^*(P_3)$. This choice, however, gives us the mode functions, which correspond to the coherent state of positive and negative frequency solutions as $|z|\rightarrow +\infty$ \cite{Anderson:2013zia}:
\begin{equation}\label{F67}
	v_{p_{\perp}}(P_3) = \frac{1}{\sqrt[4]{4(m^2+|p_{\perp}|^2)}}\left[f_{p_{\perp}}^{(0)} - i\sqrt{\frac{m^2+|p_{\perp}|^2}{2eE}} f_{p_{\perp}}^{(1)}\right],
\end{equation}
i.e. such modes violate the condition of the Hadamard behaviour that we have imposed at the beginning of this section.

Actually, one can guess the solutions (\ref{F65})--(\ref{F66}) directly from (\ref{F60}), but let us treat this expression one more time in a bit different way:
\begin{gather}
	\langle J^3_{\text{\tiny scalar}}\rangle_{\text{tree}} =  \int \frac{d^3\textbf{p}}{(2\pi)^3}P_3|f_{\textbf{p}}(t)|^2 = \nonumber \\
	= \int \frac{d^2p_{\perp}}{(2\pi)^3}\int dP_3 P_3 \left[ |\alpha|^2|D_{\nu-\frac{1}{2}}(z)|^2 + |\beta|^2|D_{\nu-\frac{1}{2}}(-z)|^2 
	+ 2 \text{Re}\left(\alpha^*\beta D^*_{\nu-\frac{1}{2}}(z) D_{\nu-\frac{1}{2}}(-z)\right) \right] = \nonumber \\
	= \int \frac{d^2p_{\perp}}{(2\pi)^3}\int dP_3 P_3 \left[ (|\alpha|^2-|\beta|^2)|D_{\nu-\frac{1}{2}}(z)|^2  + 2 \text{Re}\left(\alpha^*\beta D^*_{\nu-\frac{1}{2}}(z) D_{\nu-\frac{1}{2}}(-z)\right) \right].
\label{F68}
\end{gather}
Then, using the following properties of the parabolic cylinder functions \cite{Bateman}:
\begin{gather}
	D_{\nu}(z) = e^{-\nu \pi i}D_{\nu}(-z) + \frac{\sqrt{2\pi}}{\Gamma(-\nu)}e^{-\frac{(\nu+1)\pi i}{2}}D_{-\nu-1}(iz) = \nonumber \\
	= e^{\nu \pi i}D_{\nu}(-z) + \frac{\sqrt{2\pi}}{\Gamma(-\nu)}e^{\frac{(\nu+1)\pi i}{2}}D_{-\nu-1}(-iz),
\label{F69}
\end{gather}
one can rewrite (\ref{F60}) in the form
\begin{equation}\label{F70}
		\langle J^3_{\text{\tiny scalar}}\rangle_{\text{tree}} = \int \frac{d^2p_{\perp}}{(2\pi)^3}\int dP_3 P_3 \left[ 	|\alpha|^2 - |\beta|^2 +2\text{Im}(\alpha^*\beta)e^{\pi\frac{m^2+|p_{\perp}|^2}{2eE}} \right] |D_{\nu-\frac{1}{2}}(z)|^2.
\end{equation}
 Now one can see the condition (\ref{F62}) without any approximations. 

Returning back to fermions, in the same way we find that in this case

\begin{equation}\label{F71}
	\langle J^3\rangle_{\text{tree}} = \int \frac{d^3\textbf{p}}{(2\pi)^3} 4(m^2+|p_{\perp}|^2)\left(|\alpha|^2 + |\beta|^2 +2\text{Re}(\alpha^*\beta)e^{\pi\frac{m^2+|p_{\perp}|^2}{2eE}} \right) \left[ 1 - 2|D_{\nu}(z)|^2 e^{-\pi\frac{m^2+|p_{\perp}|^2}{4eE}} \right] e^{\pi \frac{m^2+|p_{\perp}|^2}{4eE}}.
\end{equation}
Now we explicitly see the condition (\ref{F52}) for the tree--level current to vanish. Furthermore, we observe that there should not be any finite contribution to the tree--level current is not just a coincidence, i.e. only $\sim \Lambda$ and $\sim 1/\Lambda$ and higher contributions can be present. 

\section{Conclusion}\label{Conclusion}

We study the dependence of such a physical quantity as the current of created particles on the choice of the initial state of the theory in two different non-stationary situations. In the case of pulse background we derive the expressions (\ref{termres02big})--(\ref{termres02small}) for the currents with in--vacuum and thermal density matrix as initial states at past infinity. 

The situation of constant electric field is less physically grounded. As the result we observe either vanishing or linearly divergent currents depending on the initial state. However, one can consider this case as a model example similar to the quantum field theory in de Sitter space-time \cite{Akhmedov:2013vka}, \cite{Krotov:2010ma}, \cite{Anderson:2013ila, Anderson:2013zia}. Nevertheless, we derive the expression (\ref{F49}), which establishes the dependence of the current on the choose of Fock's vacuum state, parameterized by the coefficients $\alpha, \; \beta$. We conclude that the current in this case is either zero or divergent. 

However, there is an important message, which is in order here. Specifically, there is a much bigger zoo of various choices of the mode functions and corresponding Fock space vacua. In fact, one can set the generalized Bogolyubov's coefficients instead of $\alpha, \beta$ and construct the modes as follows

\begin{equation}\label{F72}
	\xi_{\textbf{p},s}(t,\textbf{x}) = \int d^3\textbf{k} \sum\limits_{r=1}^{2} \left[ \alpha_{\textbf{kp}r}^s \psi^{\left( + \right)}_{\textbf{k},s}(t)e^{i\textbf{kx}} + \beta_{-\textbf{kp}r}^s \psi^{\left( - \right)}_{-\textbf{k},s}(t)e^{-i\textbf{kx}}  \right],
\end{equation}
\begin{equation}\label{F73}
	\eta_{\textbf{p},s}(t,\textbf{x}) = \int d^3\textbf{k} \sum\limits_{r=1}^{2} \left[ \gamma_{\textbf{kp}r}^s \psi^{\left( + \right)}_{\textbf{k},s}(t)e^{i\textbf{kx}} + \omega_{-\textbf{kp}r}^s \psi^{\left( - \right)}_{-\textbf{k},s}(t)e^{-i\textbf{kx}}  \right].
\end{equation}
The only physical restriction one should impose here is that $\beta$'s and $\gamma$'s should tend to zero while $\alpha$'s and $\omega$'s should tend to delta-functions as either of the modulus of their $\textbf{p}$ or $\textbf{k}$ arguments tends to infinity. That is necessary for the tree--level propagators to have the proper UV behaviour.

In any case the latter choice of the modes extends the space of possible states for consideration. The same is true in other eternal backgrounds, including the gravitational ones.

The main goal of our note is to make a step towards the understanding of how to treat quantum fields in the presence of strong background fields of various nature. Here we consider the problem at tree-level, but the same questions arise for loop quantum corrections. The latter issue will be studied elsewhere.

\section{Acknowledgements}

We would like to acknowledge valuable discussions with D. Trunin, F. Popov, A.Semenov and M.Visotskiy.

The work of ETA was supported by the grant from the Foundation for the Advancement of Theoretical Physics and Mathematics ``BASIS'' and by RFBR grant 18-01-00460 and by the joint
RFBR-MOST grant 21-52-52004 (E.T.). Our joint work is supported by Russian Ministry of education and science.

\appendix

\section{Klein paradox}\label{App_A}

\subsection{Modes}\label{Modes}

For completeness in this appendix we consider (1+1)-dimensional theory of fermions with electric step-potential as a classical background $eA_{\mu}(x) = V_0\theta(x)\delta_{0\mu}$, assuming that $V_0>2m$. We want to observe the effect of the Klein paradox using our methods via the current. Electric field turns out to be the thin delta-functional wall $E(x)=-V_0\delta(x)$, which separates positive and negative semi--spaces. Dirac's equation in such a background is 
	\begin{equation}\label{A1}
		\left(i\gamma^{\mu}\partial_{\mu}-e\gamma^{\mu}A_{\mu}-m\right)\psi(t,x)=0,
	\end{equation}
	where we use the following representation of gamma-matrices:
	\begin{equation}\label{A2}
		\gamma^{0} = \begin{pmatrix}
		1 & 0\\
		0 & -1
		\end{pmatrix}, \; \gamma^{1} = \begin{pmatrix}
		0 & i\\
		i & 0
		\end{pmatrix}.
	\end{equation}
	Looking for positive-frequency solutions we use the ansatz $\psi(t,x)=e^{-i\omega t}\psi_{\omega}(x) = e^{-i\omega t}\begin{pmatrix}
	\psi_L(x)\\
	\psi_R(x)
	\end{pmatrix}$ and rewrite (\ref{A1}) through its components as:
	\begin{equation}\label{A3}
		\begin{cases}
		(\omega - V_0\theta(x))\psi_L(x) - \partial_x \psi_R(x) - m\psi_L(x) = 0,\\
		(\omega - V_0\theta(x))\psi_R(x) + \partial_x \psi_L(x) + m\psi_R(x) = 0.
		\end{cases}
	\end{equation}
	One can note that physically Dirac's sea in the positive semi-space is shifted by $V_0$ and due to the condition $V_0>2m$ there is the overlapping of electron's and hole's energy levels, which is known as the Klein zone $\omega\in [m,V_0-m]$. We will see below that it is this region which determines the current of created particles. It is convenient to introduce the functions
	
	\begin{gather}
		\chi_1(\omega) = \begin{cases}
		0, \; \omega\in[-m,m],\\
		1, \; \text{otherwise}
		\end{cases},
		\;\;
		\chi_2(\omega) = \begin{cases}
		0, \; \omega\in[V_0-m,V_0+m],\\
		1, \; \text{otherwise}
		\end{cases},
		\nonumber \\
		\chi_3(\omega) = \begin{cases}
		-1, \; \omega\in[m,V_0-m],\\
		1, \; \text{otherwise}
		\end{cases}.
	\label{A4}
	\end{gather}
	Solutions of (\ref{A3}), that satisfy continuity conditions are as follows:
	
	\begin{gather}
		\psi_{\omega}^{(+)}(x) = A\left\{ \frac{2k}{p(\varkappa+1)}\begin{pmatrix}
		ip\chi_3(\omega)\\
		\omega-V_0-m
		\end{pmatrix}e^{i\chi_3(\omega)px}\theta(x) + \left[\begin{pmatrix}
		ik\\
		\omega-m
		\end{pmatrix}e^{ikx} + \frac{\varkappa-1}{\varkappa+1}\begin{pmatrix}
		-ik\\
		\omega-m
		\end{pmatrix}e^{-ikx}  \right]\theta(-x)\right\}, \\
		\psi_{\omega}^{(-)}(x) = B\bigg\{ \left[\begin{pmatrix}
		-ip\\
		\omega-V_0-m
		\end{pmatrix}e^{-ipx} + \frac{1-\varkappa}{1+\varkappa}\begin{pmatrix}
		ip\\
		\omega-V_0-m
		\end{pmatrix}e^{ipx}  \right]\theta(x) + \nonumber \\
		+ \frac{2\varkappa p}{k(\varkappa+1)}\begin{pmatrix}
		-ik\chi_3(\omega)\\
		\omega-m
		\end{pmatrix}e^{-i\chi_3(\omega)kx}\theta(-x) \bigg\},
	\label{A6}
	\end{gather}
	where
	\begin{gather}
		k=\sqrt{\omega^2-m^2}, \; p=\sqrt{(\omega-V_0)^2-m^2} \\
		\varkappa = \frac{k}{p}\frac{\omega-V_0-m}{\omega-m}\chi_3(\omega)
	\label{A8}
	\end{gather}
	and in the regions $[-m,m],\; [V_0-m,V_0+m]$ we assume that $k=i|k|,\;p=i|p|$, correspondingly, and $\varkappa=i|\varkappa|$. In all other energy regions quantities $k,\;p$ and $\varkappa$ are positive real numbers.
	
	\subsection{Field expansion}
	To count all independent solutions of Dirac's equation we expand the fermions field as follows:
	\begin{equation}\label{A9}
		\psi(t,x) = \int_{-\infty}^{+\infty}\frac{d\omega}{2\pi}\left\{e^{-i\omega t}\left[a_{\omega}\psi_{\omega}^{(+)}(x)+b_{\omega}\psi_{\omega}^{(-)}(x)\right]+e^{i\omega t}\left[a_{\omega}^{\dagger}\psi_{-\omega}^{(+)*}(x)+b_{\omega}^{\dagger}\psi_{-\omega}^{(-)*}(x)\right]\right\},
	\end{equation}
	and impose the standard anticommutation relations on the creation-annihilation operators (we write down only non-trivial ones)
	\begin{gather}
		\{a_{\omega'},a_{\omega}^{\dagger}\} = 2\pi\delta(\omega'-\omega), \; \{b_{\omega'},b_{\omega}^{\dagger}\} = 2\pi\delta(\omega'-\omega).
	\label{A10}
	\end{gather}
	In order to represent the anticommutator $\{\psi_{a}^{\dagger}(t,x),\psi_{b}(t,y)\}$ in the semi-canonical form 
	$$\{\psi_{a}^{\dagger}(t,x),\psi_{b}(t,y)\} = \delta(y-x)\delta_{ab} + I_{ab}^{\sigma_x\sigma_y}(x,y), \; \sigma_x=\text{sign}(x),$$ 
	where $I_{ab}^{\sigma_x\sigma_y}(x,y)$ is some regular at $x=y$ quantity, we should take the following normalization coefficients:
	\begin{equation}\label{A11}
		|A|^2 = \frac{1}{4}\frac{|p|}{|\omega-V_0-m|}\frac{|\varkappa|}{|k|^2}\chi_1(\omega),
	\end{equation}
	\begin{equation}\label{A12}
	|B|^2 = \frac{1}{4}\frac{|p|}{|\omega-V_0-m|}\frac{1}{|p|^2}\chi_2(\omega).
	\end{equation} 
	For instance, we have
	\begin{gather}
	I_{00}^{++} = 2\int_{V_0-m}^{V_0+m}\frac{d\omega}{2\pi}\frac{|\varkappa|}{|1+\varkappa|^2} \frac{|p|}{|\omega-V_0-m|}e^{-|p|(y+x)} - \frac{1}{2}\int_{-\infty}^{+\infty}\frac{d\omega}{2\pi}\frac{|p|}{|\omega-V_0-m|}\left\{ \frac{1-\varkappa}{1+\varkappa}e^{-ip(y+x)} + c.c. \right\}\chi_2(\omega), \\
    I_{00}^{--} = 2\int_{-m}^{m}\frac{d\omega}{2\pi}\frac{|\varkappa|}{|1+\varkappa|^2} \frac{|k|}{|\omega-m|}e^{|k|(y+x)} - \frac{1}{2}\int_{-\infty}^{+\infty}\frac{d\omega}{2\pi}\frac{|k|}{|\omega-m|}\left\{ \frac{\varkappa-1}{\varkappa+1}e^{-ik(y+x)} + c.c. \right\}\chi_1(\omega)
\label{A15}
	\end{gather}
and some similar expressions for other components. Thus, for general mass the modes under consideration do not fulfil the canonical commutation relations simultaneously for the ladder operators and fermions.

We note that in the massless case $m=0$ all the integrals above can be taken and one sees that in this case we obtain the canonical commutation relations, $\{\psi_{a}^{\dagger}(t,x),\psi_{b}(t,y)\} = \delta(y-x)\delta_{ab}$, in each (positive or negative) semispace, but the anticommutator is not equal to the delta-function if we take points $x$ and $y$ from different semi--spaces.

\subsection{Current of created particles}

We consider the vacuum $|0\rangle$ as the state annihilated by operators $a_{\omega}, b_{\omega}$ from (\ref{A9}). Then for the current one obtains:

\begin{gather}
	\langle 0|j_1(x)|0\rangle = \langle 0|\overline{\psi}(t,x) \gamma^1 \psi(t,x)|0\rangle = \int_{-\infty}^{\infty}\frac{d\omega}{2\pi}\text{sign}(\omega-V_0-m)\frac{2|\varkappa|}{|1+\varkappa|^2}\chi_1(\omega)\chi_2(\omega)\left(1-\chi_3(\omega)\right) = \nonumber \\
	= -\int_{m}^{V_0-m}\frac{d\omega}{2\pi}\frac{4\varkappa}{(1+\varkappa)^2},
\label{A16}
\end{gather}
reproducing the result from \cite{History}. As it was expected, the current of created particles is completely determined by the Klein zone. For example, in the massless case we get
\begin{equation}\label{A17}
	\langle 0|j_1(x)|0\rangle = -\frac{V_0}{2\pi},
\end{equation}
while for small mass
\begin{equation}\label{A18}
\langle 0|j_1(x)|0\rangle \simeq -\frac{V_0-2m}{2\pi}.
\end{equation}

\end{document}